\documentclass{article}

\usepackage{amsmath, amssymb, latexsym, amsfonts}
\usepackage{epsf}

\def\sect#1{\section{#1}\setcounter{equation}{0}}
\def\q{\boldsymbol{q}}
\def\p{\boldsymbol{p}}

\def\w{\boldsymbol{v}}
\def\z{\boldsymbol{z}}
\def\ww{\boldsymbol{w}}

\begin{document}


\centerline{} \vskip0.5cm \centerline{\LARGE \bf Total Variation
in Hamiltonian Formalism} \vskip3pt \centerline{\LARGE \bf and
Symplectic-Energy integrators } \vskip0.7cm \centerline{\large
Jing-Bo Chen,\quad Han-Ying Guo\quad and\quad Ke Wu} \vskip3pt
\centerline{\small Institute of Theoretical Physics, Chinese
Academy of Sciences} \vskip3pt \centerline{\small P.O. Box  2735,
Beijing 100080, P.R. China} \vskip3pt \centerline{\small
\texttt{<chenjb><hyguo><wuke>@itp.ac.cn}} \vskip1cm


\begin{abstract}
We present a discrete total variation calculus in Hamiltonian
formalism in this paper. Using this discrete variation calculus
and generating functions for the flows of Hamiltonian systems, we
derive two-step symplectic-energy integrators of any finite order
for Hamiltonian systems from a variational perspective. The
relationship between symplectic integrators derived directly from
the Hamiltonian systems and the variationally derived
symplectic-energy integrators is explored.

\end{abstract}



\sect{Introduction}

We begin by recalling the ordinary variational principle in Hamiltonian formalism.
Suppose $Q$ denotes the  configuration space with coordinates $q^{i}$, and
$T^{*}Q$ the phase space with coordinates $(q^{i}, p^{i})$, $i=1, 2,\cdots, n$.
Consider a Hamiltonian $H:\,\,T^{*}Q  \to \mathbb{R}$.
The corresponding action functional is defined by
\begin{align}
    S((q^{i}(t), p^{i}(t)))=\int_{a}^{b}(p^{i}\dot{q}^{i}-H(q^{i}, p^{i}))
                             \,dt, \label{1.1}
\end{align}
where $(q^{i}(t), p^{i}(t))$ is a $C^{2}$ curve in phase space $T^{*}Q$.

The variational principle in Hamiltonian formalism seeks the curves
$(q^{i}(t), p^{i}(t))$ for which the action functional
$S$ is stationary under variations of $(q^{i}(t), p^{i}(t))$ with fixed
endpoints.  We first define the variation of $(q^{i}(t), p^{i}(t))$.

Let
\begin{align}
   V=\sum_{i=1}^{n}\phi^{i}(\q, \p)\frac{\partial}{\partial q^{i}}
         +\sum_{i=1}^{n}\psi^{i}(\q, \p)\frac{\partial}{\partial p^{i}} \label{1.2}
\end{align}
be a  vector field on $T^{*}Q$. Here $\q=(q^{1},\cdots, q^{n}),\,\,
\p=(p^{1}, \cdots, p^{n}).$ For simplicity, we will omit the summation notation
$\sum$ in the following.

Denote the flow of $V$ by $F^{\epsilon}$:  $F^{\epsilon}(\q, \p)=
(\tilde{\q}, \tilde{\p})$, which is written in components as
\begin{align}
  &\tilde{q}^{i}=f^{i}(\epsilon, \q, \p),\label{1.3}\\
  &\tilde{p}^{i}=g^{i}(\epsilon, \q, \p),  \label{1.4}
\end{align}
where $(\q, \p)\in T^{*}Q$ and
\begin{align*}
   &\left.\frac{d}{d\epsilon}\right|_{\epsilon=0}f^{i}(\epsilon, \q, \p)
            =\phi^{i}(\q, \p).\\
   &\left.\frac{d}{d\epsilon}\right|_{\epsilon=0}g^{i}(\epsilon, \q, \p)
            =\psi^{i}(\q, \p).
\end{align*}

Let $(q^{i}(t), p^{i}(t))$ be a curve in $T^{*}Q$.  The
transformations (\ref{1.3}-\ref{1.4}) transform $(q^{i}(t),
p^{i}(t))$  into a family of curves
\[
         (\tilde{q}^{i}(t), \tilde{p}^{i}(t))
         =(f^{i}(\epsilon, \q(t), \p(t)), g^{i}(\epsilon, \q(t), \p(t))).
\]
Now we are ready to define the variation of $(q(t), p(t))$:
\begin{align}
   \delta(q^{i}(t), p^{i}(t))=:\left.\frac{d}{d\epsilon}\right|_{\epsilon=0}
      (\tilde{q}^{i}(t), \tilde{p}^{i}(t))=(\phi^{i}(\q, \p), \psi^{i}(\q, \p)).\label{1.5}
\end{align}
Next, we calculate the variation of $S$ at $(q^{i}(t), p^{i}(t))$
\small
\begin{align}
\delta S=&\left.\frac{d}{d\epsilon}\right|_{\epsilon=0}
            S((\tilde{q}^{i}(t), \tilde{p}^{i}(t))) \notag\\
        =&\left.\frac{d}{d\epsilon}\right|_{\epsilon=0}
            S((f^{i}(\epsilon, \q(t), \p(t)), g^{i}(\epsilon, \q(t), \p(t)))) \notag\\
        =&\left.\frac{d}{d\epsilon}\right|_{\epsilon=0}\int_{a}^{b}
          (g^{i}(\epsilon, \q(t), \p(t))\frac{d}{dt}f^{i}(\epsilon, \q(t), \p(t))
          -H(f^{i}(\epsilon, \q(t), \p(t)), g^{i}(\epsilon, \q(t), \p(t))))
          dt\notag\\
        =&\int_{a}^{b}\left[\left(\dot{q}^{i}-\frac{\partial H}{\partial p^{i}}\right)
          \psi^{i}+\left(-\dot{p}^{i}-\frac{\partial H}{\partial q^{i}}\right)\phi^{i} \right]dt
          +\left.p^{i}\phi^{i}\right|_{a}^{b}.\label{1.6}
\end{align}
\normalsize If $\phi^{i}(\q(a), \p(a))=\phi^{i}(\q(b), \p(b))=0$,
the requirement of $\delta S =0$ yields the Hamilton equation for
$(q^{i}(t), p^{i}(t))$
\begin{align}
\begin{split}
  & \dot{q}^{i}=\frac{\partial H}{\partial p^{i}},\\
  & \dot{p}^{i}=-\frac{\partial H}{\partial q^{i}}.
\end{split} \label{1.7}
\end{align}
If we drop the requirement of $\phi^{i}(\q(a),
\p(a))=\phi^{i}(\q(b), \p(b))=0$, we can naturally obtain the
canonical one form on $T^{*}Q$ from the second term in
(\ref{1.6}). $\theta=p^{i}dq^{i}$.  Furthermore, restricting
$(\tilde{q}^{i}(t), \tilde{p}^{i}(t))$ to the solution space of
(\ref{1.7}), we can prove the solution of (\ref{1.7}) preserves
the canonical two form $\omega=d\theta_{L}=dp^{i}\wedge dq^{i}$.

On the other hand, it is not necessary to restrict $(\tilde{q}^{i}(t),
\tilde{p}^{i}(t))$ to the solution space of (\ref{1.7}). Introducing the Euler-Lagrange
one-form
\begin{align}
           E(q^{i}, p^{i})=\left(\dot{q}-\frac{\partial H}{\partial p}\right)
          dp^{i}+\left(-\dot{p}-\frac{\partial H}{\partial q}\right)dq^{i}, \label{1.8}
\end{align}
the nilpotency of $d$ leads to
\begin{align}
  dE(q^{i}, p^{i})+\frac{d}{dt}\omega=0.\label{1.9}
\end{align}
Namely, the necessary and sufficient condition for symplectic
structure preserving is that the Euler-Lagrange one form
(\ref{1.8}) is closed [\ref{glw}, \ref{guo1}, \ref{guo2}].

Based on the above variational principle in Hamiltonian formalism and using
the ideas of discrete Lagrange mechanics
[\ref{m1}, \ref{m3}, \ref{v1}, \ref{v2}, \ref{w1}],
we can develop a natural version
of discrete Hamiltonian mechanics with fixed time steps and drive symplectic
integrators for Hamilton canonical equations from a variational perspective
[\ref{guo2}].

 However the symplectic integrators
obtained in this way are not energy-preserving in general because of its fixed
time steps [\ref{g1}, \ref{s1}].
 An energy-preserving symplectic  integrator
is a more preferable and natural candidate of approximations for
conservative Hamilton equations  since the solution of
conservative Hamilton equations is not only symplectic but also
energy-preserving. To attain this goal, we use variable time steps
and a discrete total variation calculus developed in [\ref{k1},
\ref{l1}, \ref{l2}, \ref{l3}, \ref{c1}]. The basic idea is to
construct a discrete action functional and then to apply a
discrete total variation calculus. In this way, we can derive
symplectic integrators and their associated energy conservation
laws. These variationaly derived symplectic integrators are
two-step integrators. If we take fixed time steps, the resulting
integrators are equivalent to the symplectic integrators derived
directly from the Hamiltonian systems in some special cases.

An outline of this paper is as follows. In Section 2, we
present total variation for continuous variational principle in Hamiltonian
formalism.  Section 3 is devoted to deriving symplectic-energy integrators.
In Section 4, using generating function methods, we obtain high order
symplectic-energy integrators.
 We finish this paper by making some conclusions and comments in Section 5.



\sect{Total variation in Hamiltonian formalism }

In order to discuss total variation in Hamiltonian formalism, we
will work with extended phase space $\mathbb{R}\times T^{*}Q$ with
coordinates $ (t, q^{i}, p^{i})$ [\ref{a1}]. Here $t \in
\mathbb{R}$ denotes time. 
By total
variation, we refer to variations of both $(q^{i}, p^{i})$ and
$t$. Consider a  vector field on $\mathbb{R}\times T^{*}Q$
\begin{align}
 V=\xi(t,\q, \p)\frac{\partial}{\partial t}
    +\phi^{i}(t,\q, \p)\frac{\partial}{\partial q^{i}}
     +\psi^{i}(t,\q, \p)\frac{\partial}{\partial p^{i}}.  \label{2.1}
\end{align}
\noindent
Let $F^{\epsilon}$ be the flow of $V$. For $(t,q^{i}, p^{i})\in \mathbb{R}\times T^{*}Q$,
 we have $F^{\epsilon}(t, q^{i}, p^{i})=(\tilde{t}, \tilde{q}^{i},
 \tilde{p}^{i})$, which can be written as
\begin{align}
  &\tilde{t}=h(\epsilon, t, \q, \p),\label{2.2}\\
  &\tilde{q}^{i}=f^{i}(\epsilon, t, \q, \p),  \label{2.3}\\
  &\tilde{p}^{i}=g^{i}(\epsilon, t, \q, \p),  \label{2.4}
\end{align}
where
\begin{align}
   &\left.\frac{d}{d\epsilon}\right|_{\epsilon=0}h(\epsilon, t, \q, \p)
            =\xi(t,\q, \p), \label{2.5}\\
   &\left.\frac{d}{d\epsilon}\right|_{\epsilon=0}f^{i}(\epsilon, t, \q, \p)
            =\phi^{i}(t,\q, \p), \label{2.6}\\
   &\left.\frac{d}{d\epsilon}\right|_{\epsilon=0}g^{i}(\epsilon, t, \q, \p)
            =\psi^{i}(t,\q, \p). \label{2.7}
\end{align}
 The transformation (\ref{2.5}-\ref{2.7}) transforms a curve
$(q^{i}(t), p^{i}(t))$  into a family of curves $(\tilde{q}^{i}(\epsilon,
\tilde{t}), \tilde{p}^{i}(\epsilon, \tilde{t}))$  determined by
\begin{align}
  &\tilde{t}=h(\epsilon, t, \q(t), \p(t)),\label{2.8} \\
  &\tilde{q}^{i}=f^{i}(\epsilon, t, \q(t), \p(t)).  \label{2.9}\\
  &\tilde{p}^{i}=g^{i}(\epsilon, t, \q(t), \p(t)).  \label{2.10}
\end{align}
Suppose we can solve (\ref{2.8}) for $t$: $t=h^{-1}(\epsilon,
\tilde{t})$. Then we have
\begin{align}
 &\tilde{q}^{i}(\epsilon, \tilde{t})=f^{i}(\epsilon, h^{-1}(\epsilon, \tilde{t}),
 \q(h^{-1}(\epsilon, \tilde{t})), \p(h^{-1}(\epsilon, \tilde{t}))). \label{2.11}\\
&\tilde{p}^{i}(\epsilon, \tilde{t})=g^{i}(\epsilon, h^{-1}(\epsilon, \tilde{t}),
 \q(h^{-1}(\epsilon, \tilde{t})), \p(h^{-1}(\epsilon, \tilde{t}))). \label{2.12}
\end{align}

Before calculating the variation of $S$ directly, we first consider the first order
prolongation of $V$
\begin{align}
\begin{split}
 \text{pr}^{1}V=&\xi(t, \q, \p)\frac{\partial}{\partial t}+
      \phi^{i}(t, \q, \p)\frac{\partial}{\partial q^{i}}
      + \psi^{i}(t, \q, \p)\frac{\partial}{\partial p^{i}} \\
      &\quad +\alpha^{i}(t, \q, \dot{\q}, \dot{\p})\frac{\partial}
      {\partial \dot{q}^{i}}
      +\beta^{i}(t, \q, \dot{\q}, \dot{\p})\frac{\partial}{\partial \dot{p}^{i}},
\end{split}\label{2.13}
\end{align}
where $\text{pr}^{1}V$ denote the first order prolongation
of $V$ and
\begin{align}
      &\alpha^{i}(t, \q, \dot{\q}, \dot{\p})=
       D_{t}\phi^{i}(t, \q, \p)-\dot{q}^{i}D_{t}\xi(t, \q, \p), \label{2.14}\\
      &\beta^{i}(t, \q, \dot{\q}, \dot{\p})=
       D_{t}\psi^{i}(t, \q, \p)-\dot{p}^{i}D_{t}\xi(t, \q, \p), \label{2.15}
\end{align}
where $D_{t}$ denotes the total derivative. For example
\begin{align*}
  D_{t}\phi^{i}(t, \q, \p)=\phi^{i}_{t}+\phi^{i}_{\q}\dot{\q}+\phi^{i}_{\p}\dot{\p}.
\end{align*}
For prolongation of vector field and the formula (\ref{2.14}-\ref{2.15}), we refer
the reader to [\ref{o1}].

Now we calculate the variation of $S$ directly
\begin{align}
\delta S=&\left.\frac{d}{d\epsilon}\right|_{\epsilon=0}
       S((\tilde{q}^{i}(\epsilon, \tilde{t}), \tilde{p}^{i}(\epsilon, \tilde{t}))
           \notag\\
        =&\left.\frac{d}{d\epsilon}\right|_{\epsilon=0}\int_{\tilde{a}}^{\tilde{b}}
           \left(\tilde{p}^{i}(\epsilon, \tilde{t})) \frac{d}{d\tilde{t}}
           \tilde{q}^{i}(\epsilon, \tilde{t})-H(\tilde{q}^{i}(\epsilon, \tilde{t}),
           \tilde{p}^{i}(\epsilon, \tilde{t}))\right)\,d\tilde{t}\notag\\
       =&\left.\frac{d}{d\epsilon}\right|_{\epsilon=0}\int_{a}^{b}
           \left(\tilde{p}^{i}(\epsilon, \tilde{t})) \frac{d}{d\tilde{t}}
           \tilde{q}^{i}(\epsilon, \tilde{t})-H(\tilde{q}^{i}(\epsilon, \tilde{t}),
           \tilde{p}^{i}(\epsilon, \tilde{t}))\right)\frac{d\tilde{t}}{dt}\,dt
          \quad\quad\quad (\tilde{t}=h(\epsilon, t, \q(t),\p(t)))\notag\\
        =&\int_{a}^{b}\left.\frac{d}{d\epsilon}\right|_{\epsilon=0}
           \left(\tilde{p}^{i}(\epsilon, \tilde{t})) \frac{d}{d\tilde{t}}
           \tilde{q}^{i}(\epsilon, \tilde{t})-H(\tilde{q}^{i}(\epsilon, \tilde{t}),
           \tilde{p}^{i}(\epsilon, \tilde{t})\right)\,dt\notag\\
           & \quad\quad+\int_{a}^{b}\left(p^{i}(t)\dot{q}^{i}(t)-H(q^{i}(t), p^{i}(t))
            \right) D_{t}\xi dt \label{2.16}\\
         =&\int_{a}^{b}\left[\left(\frac{d}{dt}H(q^{i}(t), p^{i}(t))\right)\xi
          +\left(-\dot{p}^{i}-\frac{\partial H}{\partial q^{i}}\right)\phi^{i}
          +\left(\dot{q}^{i}-\frac{\partial H}{\partial p^{i}}\right)
          \psi^{i}\right]dt\notag \\
          &\quad \quad
           +\left.\left[p^{i}\phi^{i}-H(q^{i}, p^{i}))\xi\right]\right|_{a}^{b}.
           \label{2.17}
\end{align}
\noindent
Here in (\ref{2.16}), we used (\ref{2.5}) and the fact
 \[
  \left.\frac{d}{d\epsilon}\right|_{\epsilon=0}
  \frac{d\tilde{t}}{dt}=\frac{d}{dt}\left.\frac{d}{d\epsilon}\right|_{\epsilon=0}
 \tilde{t}=D_{t}\xi.
\]
 In (\ref{2.17}), we used the prolongation formula (\ref{2.14}).

If $\xi(a, \q(a), \p(a))=\xi(b, \q(b), \p(a))=0$ and  
$\phi^{i}(a, \q(a), \p(a))=\phi^{i}(b, \q(b), \p(b))=0$,
 the requirement of $\delta S
=0$ yields the Hamilton canonical equation
\begin{align}
\begin{split}
  & \dot{q}^{i}=\frac{\partial H}{\partial p^{i}},\\
  & \dot{p}^{i}=-\frac{\partial H}{\partial q^{i}},
\end{split} \label{2.18}
\end{align}
from the  variation $\phi^{i}, \psi^{i}$ and
the energy conservation law
\begin{align}
 \frac{d}{dt}H(q^{i}, p^{i})=0 \label{2.19}
\end{align}
from the variation  $\xi$.

 Since
\begin{align*}
  \frac{d}{dt}H(q^{i}, p^{i})=\frac{\partial H}{\partial q^{i}}\dot{q}^{i}+
                      \frac{\partial H}{\partial p^{i}}\dot{p}^{i},
\end{align*}
 we can easily see that the energy conservation law (\ref{2.19}) is a natural
consequence of the Hamilton canonical equations (\ref{2.18}).

If we drop the requirement
\begin{align*}
 &\xi(a, \q(a), \p(a))=\xi(b, \q(b), \p(b))=0,\\
 &\phi^{i}(a, \q(a), \p(a))=\phi^{i}(b, \q(b), \p(b))=0,
\end{align*}
we can define the extended canonical one form on $\mathbb{R}\times T^{*}Q$ from
the second term in (\ref{2.17})
\begin{align}
\theta=p^{i}dq^{i}-H(q^{i}, p^{i})dt. \label{2.20}
\end{align}
Furthermore, restricting
$(\tilde{q}^{i}(t), \tilde{p}^{i}(t))$ to the solution space of (\ref{2.18}), we can
prove the solution of (\ref{2.18}) preserves
the extended canonical two form
\begin{align}
   \omega=d\theta=dp^{i}\wedge dq^{i}-dH(q^{i}, p^{i})\wedge dt, \label{2.21}
\end{align}
by using the same method in [\ref{m1}]. \vskip6pt \noindent {\em
Remark:} The vector field (\ref{2.1}) can be decomposed along
$(q^{i}(t), p^{i}(t))$ into vertical and horizontal components.
The vertical component is
\[
 (\phi^{i}-\xi \dot{q}^{i})\frac{\partial}{\partial q^{i}}
 +(\psi^{i}-\xi \dot{p}^{i})\frac{\partial}{\partial p^{i}}
 \]
 and the horizontal component is
 \[
   \xi\frac{\partial}{\partial t}+\xi \dot{q}^{i}
   \frac{\partial}{\partial q^{i}}+\xi \dot{p}^{i}
   \frac{\partial}{\partial p^{i}}.
 \]
 Then analogous to the Lagrangian formalism in [\ref{c1}], the total variation gives rise to the Hamilton equations along the vertical
 direction as well as a relation between the Hamilton equations and conservation
 law for the Hamiltonian along the horizontal direction (see also [\ref{m1}]).


\newpage
\sect{A discrete total variation calculus in Hamiltonian formalism
      and symplectic-energy integrators}

In this section, we develop a discrete version of total variation
in Hamiltonian formalism. Using this discrete total variation
calculus, we will derive symplectic-energy integrators.

 Let
\[
            L(q^{i}, p^{i}, \dot{q}^{i}, \dot{p}^{i})=p^{i}\dot{q}^{i}
                     -H(q^{i}, p^{i}),
\]
be a function from $\mathbb{R}\times T(T^{*}Q)$ to $\mathbb{R}$.
Here $L$  does not depend on $t$ explicitly.

We use $P\times P$ for the discrete version  of $\mathbb{R}\times
T(T^{*}Q)$. Here $P$ is the discrete version of $\mathbb{R}\times
T^{*}Q$. A point $(t_{0}, \q_{0}, \p_{0}; t_{1}, \q_{1},
\p_{1})\in P\times P$ corresponds to a tangent vector
$(\frac{\q_{1}-\q_{0}}{t_{1}-t_{0}}, \frac{\p_{1}-\p_{0}}
{t_{1}-t_{0}})$. For simplicity, the vector symbols $\q=(q^{1},
\cdots, q^{n})$ and $\p=(p^{1}, \cdots, p^{n})$ are used
throughout this section. A discrete $L$ is defined to be
$\mathbb{L}:\,P\times P\to \mathbb{R}$ and the corresponding
action to be
\begin{align}
 \mathbb{S}=\sum_{k=0}^{N-1}\mathbb{L}(t_{k}, \q_{k}, \p_{k}, t_{k+1}, \q_{k+1},
             \p_{k+1})(t_{k+1}-t_{k}). \label{3.1}
\end{align}

The discrete variational principle in total variation  is to
extremize $\mathbb{S}$ for variations of both $\q_{k}, \p_{k}$ and
$t_{k}$ holding the endpoints $(t_{0}, \q_{0}, \p_{0})$ and
$(t_{N}, \q_{N}, \p_{N})$ fixed. This discrete variational
principle  determines a discrete flow $\Phi:\,P\times P\to P\times
P$ by
\begin{align}
   \Phi(t_{k-1}, \q_{k-1}, \p_{k-1},  t_{k}, \q_{k}, \p_{k})
        =(t_{k}, \q_{k}, \p_{k}, t_{k+1}, \q_{k+1}, \p_{k+1}). \label{3.2}
\end{align}
Here $(t_{k+1}, \q_{k+1}, \p_{k+1})$ for all $k \in \{1, 2,
\cdots, N-1\}$ are found from the following discrete Hamilton
canonical equation  and the discrete energy conservation law
\footnotesize
\begin{align}
 \begin{split}
 &(t_{k+1}-t_{k})D_{2}\mathbb{L}(t_{k}, \q_{k}, \p_{k}, t_{k+1}, \q_{k+1}, \p_{k+1})
 +(t_{k}-t_{k-1})D_{5}\mathbb{L}(t_{k-1}, \q_{k-1}, \p_{k-1}, t_{k}, \q_{k}, \p_{k})=0,\\
&(t_{k+1}-t_{k})D_{3}\mathbb{L}(t_{k}, \q_{k}, \p_{k}, t_{k+1}, \q_{k+1}, \p_{k+1})
 +(t_{k}-t_{k-1})D_{6}\mathbb{L}(t_{k-1}, \q_{k-1}, \p_{k-1}, t_{k}, \q_{k}, \p_{k})=0,
  \end{split}
 \label{3.3}
\end{align}
\normalsize
and
\small
\begin{align}
 \begin{split}
 &(t_{k+1}-t_{k})D_{1}\mathbb{L}(t_{k}, \q_{k}, \p_{k}, t_{k+1}, \q_{k+1}, \p_{k+1})
   +(t_{k}-t_{k-1})D_{4}\mathbb{L}(t_{k-1}, \q_{k-1}, \p_{k-1}, t_{k}, \q_{k}, \p_{k}) \\
 &\quad -\mathbb{L}(t_{k}, \q_{k}, \p_{k}, t_{k+1}, \q_{k+1}, \p_{k+1})+
\mathbb{L}(t_{k-1}, \q_{k-1}, \p_{k-1}, t_{k}, \q_{k}, \p_{k})=0.
\end{split}\label{3.4}
\end{align}
\normalsize
 Here $D_{i}$ denotes the partial derivative of
$\mathbb{L}$ with respect to the $i$th argument. The Eq.
(\ref{3.3}) is the discrete Hamilton canonical  equation that is 
also called variational integrator. The Eq.(\ref{3.4}) is the discrete
energy conservation law associated to (\ref{3.3}). Unlike the
continuous case, the variational integrator (\ref{3.3}) does not
satisfy (\ref{3.4}) in general. Therefore, we need to solve
(\ref{3.3}) and (\ref{3.4}) simultaneously.

Now we prove that the discrete flow  determined by (\ref{3.3}) and
(\ref{3.4})
 preserves a discrete version of the
extended canonical two form $\omega$ defined in (\ref{2.21}). Therefore, we call  (\ref{3.3})-(\ref{3.4})
 a symplectic-energy integrator. We do this directly
from the variational point of view, consistent with the continuous case [\ref{m1}].

As in continuous case, we calculate $d\mathbb{S}$ for variations with varied
endpoints.
\begin{align}
 &d\mathbb{S}(t_{0}, \q_{0}, \p_{0} \cdots, t_{N}, \q_{N}, \p_{N})\cdot
         (\delta t_{0}, \delta \q_{0}, \delta \p_{0}
         \cdots, \delta t_{N}, \delta \q_{N}, \delta \p_{N})\notag \\
  &=\sum_{k=0}^{N-1}(D_{2}L(\w_{k})\delta \q_{k}+
        D_{5}L(\w_{k})\delta \q_{k+1}+D_{3}L(\w_{k})\delta \p_{k}+
        D_{6}L(\w_{k})\delta p_{k+1})(t_{k+1}-t_{k})\notag \\
     &\quad + \sum_{k=0}^{N-1}(D_{1}L(\w_{k})\delta t_{k}+
        D_{4}L(\w_{k})\delta t_{k+1})(t_{k+1}-t_{k})+\sum_{k=0}^{N-1}L(\w_{k})(\delta t_{k+1}
       -\delta t_{k})\notag\\
   &=\sum_{k=1}^{N-1}(D_{2}L(\w_{k})(t_{k+1}-t_{k})
                     +D_{5}L(\w_{k-1})(t_{k}-t_{k-1}))\delta \q_{k}\notag\\
                    &\quad +\sum_{k=1}^{N-1}
                     (D_{3}L(\w_{k})(t_{k+1}-t_{k})
                     +D_{6}L(\w_{k-1})(t_{k}-t_{k-1}))\delta \p_{k}\notag\\
    &\quad +\sum_{k=1}^{N-1}(D_{1}L(\w_{k})(t_{k+1}-t_{k})
                     +D_{4}L(\w_{k-1})(t_{k}-t_{k-1})+L(\w_{k-1})-L(\w_{k})
                      )\delta t_{k}\notag\\
    &\quad +D_{2}L(\w_{0})(t_{1}-t_{0})\delta \q_{0}+D_{3}L(\w_{0})
                    (t_{1}-t_{0})\delta \p_{0}
            +(D_{1}L(\w_{0})(t_{1}-t_{0})-L(\w_{0}))\delta t_{0}\notag\\
      &\quad +D_{5}L(\w_{N-1})(t_{N}-t_{N-1})\delta \q_{N}+D_{6}L(\w_{N-1})
               (t_{N}-t_{N-1})\delta \p_{N}\notag\\
       &\quad +(D_{4}L(\w_{N-1})(t_{N}-t_{N-1})-L(\w_{N-1}))\delta t_{N},  \label{3.5}
\end{align}
where $\w_{k}=(t_{k}, \q_{k}, \p_{k}, t_{k+1}, \q_{k+1},
\p_{k+1}), \quad k=0, 1, \cdots, N-1$. We can see that the last
six terms in (\ref{3.5}) come from the boundary variations. Based
on the boundary variations, we can define two one forms on
$P\times P$
\begin{align}
\begin{split}
&\theta_{\mathbb{L}}^{-}(\w_{k})
           =D_{2}\mathbb{L}(\w_{k})(t_{k+1}-t_{k}) d\q_{k}
             +D_{3}\mathbb{L}(\w_{k})(t_{k+1}-t_{k}) d\p_{k}\\
      &\, \quad +(D_{1}\mathbb{L}(\w_{k})(t_{k+1}-t_{k})
            -\mathbb{L}(\w_{k}))dt_{k} \label{3.6}
\end{split}
\end{align}
and
\begin{align}
\begin{split}
 &\theta_{\mathbb{L}}^{+}(\w_{k})
           =D_{5}\mathbb{L}(\w_{k})(t_{k+1}-t_{k}) d\q_{k+1}
             +D_{6}\mathbb{L}(\w_{k})(t_{k+1}-t_{k}) d\p_{k+1}\\
      &\, \quad +(D_{4}\mathbb{L}(\w_{k})(t_{k+1}-t_{k})
            -\mathbb{L}(\w_{k}))dt_{k+1} \label{3.7}
\end{split}
\end{align}
Here we have used the notation in [\ref{m1}]. We regard the pair $(\theta_{\mathbb{L}}^{-},
\theta_{\mathbb{L}}^{+})$ as being the discrete version of the extended canonical
one form $\theta$ defined in (\ref{2.20}).

Now we parameterize the solutions of the discrete variational principle by
$(t_{0}, q_{0}, t_{1}, q_{1})$, and  restrict $\mathbb{S}$ to that solution
space.  Then Eq. (\ref{3.5}) becomes
\begin{align}
\begin{split}
 d&\mathbb{S}(t_{0}, \q_{0}, \p_{0}, \cdots, t_{N}, \q_{N}, \p_{N})\cdot
               (\delta t_{0}, \delta \q_{0}, \delta \p_{0}\cdots, \delta t_{N},
               \delta q_{N}, \delta \p_{N})\\
 &=\theta_{\mathbb{L}}^{-}(t_{0}, \q_{0}, \p_{0}, t_{1}, \q_{1}, \p_{1})\cdot
                 (\delta t_{0}, \delta \q_{0},  \delta \p_{0}, \delta t_{1},
                 \delta \q_{1}, \delta \p_{1})\\
 &\quad +\theta_{\mathbb{L}}^{+}(t_{N-1}, \q_{N-1}, \p_{N-1},
         t_{N}, \q_{N}, \p_{N})\cdot
         (\delta t_{N-1}, \delta \q_{N-1},  \delta \p_{N-1}, \delta t_{N},
         \delta \q_{N}, \delta \p_{N},)\\
 &=\theta_{\mathbb{L}}^{-}(t_{0}, \q_{0}, \p_{0}, t_{1}, \q_{1}, \p_{1})\cdot
    (\delta t_{0}, \delta \q_{0},  \delta \p_{0}, \delta t_{1},
                 \delta \q_{1}, \delta \p_{1})\\
   &\quad +(\Phi^{N-1})^{*}\theta_{\mathbb{L}}^{+}(t_{0}, \q_{0}, \p_{0}, t_{1},
              \q_{1}, \p_{1})\cdot
    (\delta t_{0}, \delta \q_{0},  \delta \p_{0}, \delta t_{1},
                 \delta \q_{1}, \delta \p_{1}). \label{3.8}
\end{split}
\end{align}
From (\ref{3.8}), we can obtain
\begin{align}
d\mathbb{S}=\theta_{\mathbb{L}}^{-}+(\Phi^{N-1})^{*}\theta_{\mathbb{L}}^{+}.
\label{3.9}
\end{align}
The Eq. (\ref{3.9}) holds for arbitrary $N>1$. Taking N=2 leads to
\begin{align}
d\mathbb{S}=\theta_{\mathbb{L}}^{-}+\Phi^{*}\theta_{\mathbb{L}}^{+}.
\label{3.10}
\end{align}

Taking exterior differentiation of (\ref{3.10}) reveals that
\begin{align}
  \Phi^{*}(d\theta_{\mathbb{L}}^{+})=-d\theta_{\mathbb{L}}^{-}. \label{3.11}
\end{align}
From the definition of $\theta_{\mathbb{L}}^{-}$ and
$\theta_{\mathbb{L}}^{+}$, we know that
\begin{align}
   \theta_{\mathbb{L}}^{-}+\theta_{\mathbb{L}}^{+}=d\mathbb{L}.\label{3.12}
\end{align}
Taking exterior differentiation of (\ref{3.12}), we obtain
$d\theta_{\mathbb{L}}^{+}=-d\theta_{\mathbb{L}}^{-}$.  Define
\begin{align}
 \omega_{\mathbb{L}}\equiv d\theta_{\mathbb{L}}^{+}=
-d\theta_{\mathbb{L}}^{-}.\label{3.13}
\end{align}
Finally, we have shown that the discrete flow $\Phi$ preserves the discrete
extended canonical two form $\omega_{\mathbb{L}}$.
\begin{align}
  \Phi^{*}(\omega_{\mathbb{L}})=\omega_{\mathbb{L}}. \label{3.14}
\end{align}

Thus we may call the coupled difference system
(\ref{3.3})-(\ref{3.4}) symplectic-energy integrator in the sense
that it satisfies the discrete energy conservation law (\ref{3.4})
and preserves the discrete extended canonical two form
$\omega_{\mathbb{L}}$.

To illustrate the above discrete total variation calculus, we now present
an example. We choose $\mathbb{L}$ in (\ref{3.1}) as
\begin{align}
  \mathbb{L}(t_{k}, \q_{k}, \p_{k}, t_{k+1}, \q_{k+1}, \p_{k+1})
     =\p_{k+\frac{1}{2}}\frac{\q_{k+1}-\q_{k}}{t_{k+1}-t_{k}}-H(\q_{k+\frac{1}{2}},
     \p_{k+\frac{1}{2}}), \label{3.15}
\end{align}
where $\p_{k+\frac{1}{2}}=\frac{\p_{k}+\p_{k+1}}{2},~
\q_{k+\frac{1}{2}}=\frac{\q_{k}+\q_{k+1}}{2}$.

Using (\ref{3.3}), we can obtain the corresponding discrete
Hamilton equation \small
\begin{align}
\begin{split}
   &\frac{\q_{k+1}-\q_{k-1}}{2}-\frac{1}{2}\left((t_{k+1}-t_{k})\frac{\partial H}
    {\partial \p}(\q_{k+\frac{1}{2}},
    \p_{k+\frac{1}{2}})+(t_{k}-t_{k-1})\frac{\partial H}{\partial \p}(\q_{k-\frac{1}{2}},
    \p_{k-\frac{1}{2}})\right)=0,\\
  &\frac{\p_{k+1}-\p_{k-1}}{2}+\frac{1}{2}\left((t_{k+1}-t_{k})\frac{\partial H}{\partial \q}
    (\q_{k+\frac{1}{2}},
    \p_{k+\frac{1}{2}})+(t_{k}-t_{k-1})\frac{\partial H}{\partial \q}(\q_{k-\frac{1}{2}},
    \p_{k-\frac{1}{2}})\right)=0,
\end{split}\label{3.16}
\end{align}
\normalsize where $\p_{k-\frac{1}{2}}=\frac{\p_{k}+\p_{k-1}}{2},~
       \q_{k-\frac{1}{2}}=\frac{\q_{k}+\q_{k-1}}{2}$.
Using (\ref{3.4}), we can obtain the corresponding discrete energy conservation
law
\begin{align}
  H(\q_{k+\frac{1}{2}}, \p_{k+\frac{1}{2}})= H(\q_{k-\frac{1}{2}}, \p_{k-\frac{1}{2}}).
  \label{3.17}
\end{align}
The symplectic-energy integrator (\ref{3.16})-(\ref{3.17})
preserves the discrete two form
$d\theta_{\mathbb{L}}^{+}=-d\theta_{\mathbb{L}}^{-}$:
\begin{align}
\frac{1}{2}(d\p_{k}\wedge d\q_{k+1}+d\p_{k+1}\wedge d\q_{k})
  -H(\q_{k+\frac{1}{2}}, \p_{k+\frac{1}{2}})\wedge \left(\frac{dt_{k}+dt_{k+1}}{2}
  \right). \label{3.18}
\end{align}
If we take fixed time steps $t_{k+1}-t_{k}=h$ ($h$ is a constant), then (\ref{3.16})
becomes
\begin{align}
\begin{split}
   &\frac{\q_{k+1}-\q_{k-1}}{2h}=\frac{1}{2}\left(\frac{\partial H}{\partial \p}
        (\q_{k+\frac{1}{2}},
    \p_{k+\frac{1}{2}})+\frac{\partial H}{\partial \p}(\q_{k-\frac{1}{2}},
    \p_{k-\frac{1}{2}})\right),\\
  &\frac{\p_{k+1}-\p_{k-1}}{2h}=-\frac{1}{2}\left(\frac{\partial H}{\partial \q}
  (\q_{k+\frac{1}{2}},
    \p_{k+\frac{1}{2}})+\frac{\partial H}{\partial \q}(\q_{k-\frac{1}{2}},
    \p_{k-\frac{1}{2}})\right).
\end{split}\label{3.19}
\end{align}

Now we explore the relationship between (\ref{3.19}) and the mid-point integrator
for the Hamiltonian system
\begin{align}
\begin{split}
     &\dot{\q}=\frac{\partial H}{\partial \p},\\
     &\dot{\p}=-\frac{\partial H}{\partial \q}. \label{3.20}
\end{split}
\end{align}
The mid-point symplectic integrator for (\ref{3.20}) is
\begin{align}
\begin{split}
   &\frac{\q_{k+1}-\q_{k}}{h}=\frac{\partial H}{\partial \p}
        (\q_{k+\frac{1}{2}}, \p_{k+\frac{1}{2}}),\\
  &\frac{\q_{k+1}-\q_{k}}{h}=-\frac{\partial H}{\partial \q}
        (\q_{k+\frac{1}{2}}, \p_{k+\frac{1}{2}}).
\end{split}\label{3.21}
\end{align}
In (\ref{3.21}), we replace $k$ by $k-1$ and obtain
\begin{align}
\begin{split}
   &\frac{\q_{k}-\q_{k-1}}{h}=\frac{\partial H}{\partial \p}
        (\q_{k-\frac{1}{2}}, \p_{k-\frac{1}{2}}),\\
  &\frac{\p_{k}-\p_{k-1}}{h}=-\frac{\partial H}{\partial \q}
        (\q_{k-\frac{1}{2}}, \p_{k-\frac{1}{2}}).
\end{split}\label{3.22}
\end{align}
Adding (\ref{3.22}) to (\ref{3.21}) results in (\ref{3.19}). Therefore, if we use
 (\ref{3.21}) to obtain $\p_{k}, \q_{k}$, the two-step integrator (\ref{3.19}) is equivalent
to the mid-point integrator (\ref{3.21}). However, the equivalence does not
hold in general. For example,  choose $\mathbb{L}$ in (\ref{3.1}) as
\begin{align}
  \mathbb{L}(t_{k}, \q_{k}, \p_{k}, t_{k+1}, \q_{k+1}, \p_{k+1})
     =\p_{k}\frac{\q_{k+1}-\q_{k}}{t_{k+1}-t_{k}}-H(\q_{k+\frac{1}{2}},
     \p_{k+\frac{1}{2}}), \label{3.23}
\end{align}
 and take fixed time steps $t_{k+1}-t_{k}=h$. Then (\ref{3.3}) becomes
\begin{align}
\begin{split}
   &\frac{\q_{k+1}-\q_{k}}{h}=\frac{1}{2}\left(\frac{\partial H}{\partial \p}
     (\q_{k+\frac{1}{2}},
    \p_{k+\frac{1}{2}})+\frac{\partial H}{\partial \p}(\q_{k-\frac{1}{2}},
    \p_{k-\frac{1}{2}})\right),\\
  &\frac{\p_{k}-\p_{k-1}}{h}=-\frac{1}{2}\left(\frac{\partial H}{\partial \q}
        (\q_{k+\frac{1}{2}},
    \p_{k+\frac{1}{2}})+\frac{\partial H}{\partial \q}(\q_{k-\frac{1}{2}},
    \p_{k-\frac{1}{2}})\right).
\end{split}\label{3.24}
\end{align}
The integrator (\ref{3.24}) is a two-step integrator which
preserves the two form $dp_{k}\wedge dq_{k+1}$. In this case, we
cannot find one-step integrator that is equivalent to
(\ref{3.24}).

In conclusion, using discrete total variation calculus, we derive
two-step symplectic-energy integrators. When taking fixed time
steps, some of them are equivalent to one-step integrator derived
directly from the Hamiltonian system while the others do not have
this equivalence.



\sect{High order symplectic-energy integrators by generating functions}

In this section, we develop high order symplectic-energy integrators by
using the generating function of the flow of the Hamiltonian system
\begin{align}
 \dot{\z}=J\bigtriangledown H(\z), \label{4.1}
\end{align}
where $\z=(\p, \q)^{T}, J=\left(\begin{matrix}0&-I\\I&0\end{matrix}\right)$.

We first recall the generating function with normal Darboux matrix of a symplectic
transformation . For details, see [\ref{f1}, \ref{f2}].

Suppose $\alpha$ is a $4n\times 4n$ nonsingular matrix with the form
\[
    \alpha=\left(\begin{matrix}A&B\\C&D\end{matrix}\right),
\]
where $ A, B, C $ and $D$ are both $2n\times 2n$ matrices.

We denote the inverse of $\alpha$ by
\[
    \alpha^{-1}=\left(\begin{matrix}A_{1}&B_{1}\\C_{1}&D_{1}\end{matrix}\right),
\]
where $ A_{1}, B_{1}, C_{1} $ and $D_{1}$ are both $2n\times 2n$ matrices.

We call a $4n\times 4n$ matrix $\alpha$ {\em  a Darboux matrix} if
\begin{align}
   \alpha^{T}J_{4n}\alpha=\tilde{J}_{4n}, \label{4.2}
\end{align}
where
\[
    J_{4n}=\left(\begin{matrix}0&-I_{2n}\\I_{2n}&0\end{matrix}\right),\,\,
   \tilde{J}_{4n}=\left(\begin{matrix}J_{2n}&0\\0&-J_{2n}\end{matrix}\right),\,\,
   J_{2n}=\left(\begin{matrix}0&-I_{n}\\I_{n}&0\end{matrix}\right),
\]
where $I_{n}$ is an $n\times n$ identity matrix and $I_{2n}$ is a $2n\times 2n$
identity matrix.

Every Darboux matrix induces a {\em fractional transform } between symplectic and
symmetric matrices
\begin{align*}
 \sigma_{\alpha}:\quad & Sp(2n)\to Sm(2n),\\
  &\sigma_{\alpha}(S)=(AS+B)(CS+D)^{-1}=M, \quad \text{for}\quad S \in Sp(2n),
  \,\, \text{det}(CS+D)\ne0
\end{align*}
with the inverse transform $\sigma_{\alpha}^{-1}=\sigma_{\alpha_{-1}}$
\begin{align*}
 \sigma_{\alpha}^{-1}:\quad  &Sm(2n)\to Sp(2n),\\
  &\sigma_{\alpha}^{-1}(M)=(A_{1}M+B_{1})(C_{1}M+D_{1})^{-1}=S,
\end{align*}
where $Sp(2n)$ is the group of symplectic matrices and $Sm(2n)$ the set of symmetric
matrices.

We can generalize the above discussions to generally nonlinear
transformations on $\mathbb{R}^{2n}$. Denote $Spnl(2n)$ the set of
symplectic transformations on $\mathbb{R}^{2n}$ and $Smnl(2n)$ the
set of  symmetric transformations, i.e. transformations with
symmetric Jacobian, on $\mathbb{R}^{2n}$. Every $f\in Smnl(2n)$
corresponds, at least locally, to a real function $\phi$ (unique
to a constant) such that $f$ is the gradient of $\phi$:
\begin{align}
      f(\ww)=\bigtriangledown \phi(\ww), \label{4.3}
\end{align}
where $\bigtriangledown \phi(\ww)=(\phi_{w_{1}}(\ww), \cdots, \phi_{w_{2n}}(\ww))$.

Then we have
 \begin{align*}
 \sigma_{\alpha}:\quad & Spnl(2n)\to Smnl(2n),\\
  &\sigma_{\alpha}(g)=(A\circ g+B)\circ (C\circ g+D)^{-1}=
  \bigtriangledown \phi,
\end{align*}
for $g \in Spnl(2n),
  \,\, \text{det}(Cg_{\z}+D) \ne 0$, or alternatively
\begin{align*}
  Ag(\z)+B\z=(\bigtriangledown \phi)(Cg(\z)+D\z),
\end{align*}
where $\circ$ denotes the composition of transformation and the $2n\times 2n$
constant matrices $A, B, C$ and $D$ are regarded as linear transformations.
$g_{\z}$ denotes the Jacobian of the symplectic transformation $g$.

We call $\phi$ the {\em generating function} of Darboux type $\alpha$ for the
symplectic transformation $g$.

Conversely, we have
\begin{align*}
 &\sigma_{\alpha}^{-1}:\quad  Smnl(2n)\to Spnl(2n),\\
  &\sigma_{\alpha}^{-1}(\bigtriangledown \phi)=(A_{1}\circ \bigtriangledown \phi
  +B_{1})\circ (C_{1}\circ \bigtriangledown \phi+D_{1})^{-1}=g, 
\end{align*} $\text{for} \,\, \text{det}(C_{1}\phi_{\ww\ww}+D_{1}) \ne 0$,
or alternatively
\begin{align*}
 A_{1}\bigtriangledown \phi(\ww)+B_{1}\ww=g(C_{1}\bigtriangledown \phi(\ww)
  +D_{1}\ww),
\end{align*}
where $g$ is called the symplectic transformation of Darboux type $\alpha$ for the
generating function $\phi$.

For the study of integrators, we may restrict ourselves to the {\em normal
Darboux matrices}, i.e., those satisfying $A+B=0, C+D=I_{2n}$. The normal Darboux
matrices can be characterized as
\begin{align}
\alpha=\left(\begin{matrix}J_{2n}&-J_{2n}\\ E& I_{2n}-E\end{matrix}\right), \quad
E=\frac{1}{2}(I_{2n}+J_{2n}F), \quad F^{T}=F, \label{4.4}
\end{align}
and
\begin{align}
\alpha^{-1}=\left(\begin{matrix}(E-I_{2n})J_{2n}&I_{2n}\\ EJ_{2n}& I_{2n}
  \end{matrix}\right). \label{4.5}
\end{align}

The fractional transform induced by a normal Darboux matrix establishes a one-one
correspondence between symplectic transformations near {\em identity} and
symmetric transformations near {\em nullity}.

For simplicity, we take $F=0$, then $E=\frac{1}{2}I_{2n}$ and
\begin{align}
    \alpha=\left(\begin{matrix}J_{2n}&-J_{2n}\\ \frac{1}{2}I_{2n} &
   \frac{1}{2}I_{2n}\end{matrix}\right). \label{4.6}
\end{align}

Now we consider the generating function of the flow of (\ref{4.1}).  Denote
the flow of (\ref{4.1}) by $e^{t}_{H}$. The generating function $\phi(\ww,t)$ for
the flow   $e^{t}_{H}$ of Darboux type (\ref{4.6})  is give by
\begin{align}
  \bigtriangledown \phi=(J_{2n}\circ e^{t}_{H}-J_{2n})\circ \left(\frac{1}{2}
  e^{t}_{H}+\frac{1}{2}I_{2n}\right)^{-1}, \quad \text{for small} \quad |t|. \label{4.7}
\end{align}
$\phi(\ww, t)$ satisfies the Hamilton-Jacobi equation
\begin{align}
  \frac{\partial}{\partial t}\phi(\ww, t)=-H\left(\ww+\frac{1}{2}J_{2n}
   \bigtriangledown \phi(\ww, t)\right) \label{4.8}
\end{align}
and can be expressed by Taylor series in $t$
\begin{align}
 \phi(\ww, t)=\sum_{k=1}^{\infty}\phi^{k}(w)t^{k}, \quad \text{for small} \quad |t|.
 \label{4.9}
\end{align}
The coefficients $\phi^{k}(w)$ can be determined recursively
\begin{align}
\begin{split}
  &\phi^{1}(\ww)=-H(w),\\
  &\phi^{k+1}(\ww)=\frac{-1}{k+1}\sum_{m=1}^{k}\frac{1}{m!}\sum_{\underset
  {j_{l}\geqslant 1}{j_{1}+\cdots+j_{m}=k}}D^{m}H\left(\frac{1}{2}J_{2n}
  \bigtriangledown \phi^{j_{1}}, \cdots, \frac{1}{2}J_{2n}\bigtriangledown \phi
  ^{j_{m}}\right),
\end{split}\label{4.10}
\end{align}
where $ k\geqslant 1,$ and we use the notation of the $m$-linear form
\begin{align*}
  &D^{m}H\left(\frac{1}{2}J_{2n}
  \bigtriangledown \phi^{j_{1}}, \cdots, \frac{1}{2}J_{2n}\bigtriangledown \phi
  ^{j_{m}}\right) \\
  &\quad :=\sum_{i_{1}, \cdots, i_{m}=1}^{2n}H_{\z_{i_{1}}\cdots \z_{i_{m}}}
  (\ww)\left(\frac{1}{2}J_{2n}\bigtriangledown \phi^{j_{1}}(\ww)\right)_{i_{1}}\cdots
  \left(\frac{1}{2}J_{2n}\bigtriangledown \phi^{j_{m}}(\ww)\right)_{i_{m}}.
\end{align*}
From (\ref{4.7}), we can see that the phase flow $\hat{\z}:=e^{t}_{H}\z$ satisfies
\begin{align}
 J_{2n}(\hat{\z}-\z)=\bigtriangledown \phi\left(\frac{\hat{\z}-\z}{2}\right)=
     \sum_{j=1}^{\infty}t^{j}\bigtriangledown \phi^{j}
     \left(\frac{\hat{\z}+\z}{2}\right). \label{4.11}
\end{align}

Now we choose $\mathbb{L}$ in (\ref{3.1}) as
\begin{align}
  \mathbb{L}(t_{k}, \q_{k}, \p_{k}, t_{k+1}, \q_{k+1}, \p_{k+1})
     =\p_{k+\frac{1}{2}}\frac{\q_{k+1}-\q_{k}}{t_{k+1}-t_{k}}-\psi^{m}(\q_{k+\frac{1}{2}},
     \p_{k+\frac{1}{2}}), \label{4.12}
\end{align}
where
\begin{align}
 \psi^{m}(\q_{k+\frac{1}{2}}, \p_{k+\frac{1}{2}})
  =\sum_{j=1}^{m}t^{j}\phi^{j}(\q_{k+\frac{1}{2}}, \p_{k+\frac{1}{2}}).
\label{4.13}
\end{align}
The corresponding symplectic-energy integrator is
\small
\begin{align}
\begin{split}
   &\frac{\q_{k+1}-\q_{k-1}}{2}-\frac{1}{2}\left((t_{k+1}-t_{k})\frac{\partial \psi^{m}}
    {\partial \p}(\q_{k+\frac{1}{2}},
    \p_{k+\frac{1}{2}})+(t_{k}-t_{k-1})\frac{\partial \psi^{m}}{\partial \p}(\q_{k-\frac{1}{2}},
    \p_{k-\frac{1}{2}})\right)=0,\\
  &\frac{\p_{k+1}-\p_{k-1}}{2}+\frac{1}{2}\left((t_{k+1}-t_{k})\frac{\partial \psi^{m}}{\partial \q}
    (\q_{k+\frac{1}{2}},
    \p_{k+\frac{1}{2}})+(t_{k}-t_{k-1})\frac{\partial \psi^{m}}{\partial \q}(\q_{k-\frac{1}{2}},
    \p_{k-\frac{1}{2}})\right)=0,
   \\
   &\psi^{m}(\q_{k+\frac{1}{2}}, \p_{k+\frac{1}{2}})=
     \psi^{m}(\q_{k-\frac{1}{2}}, \p_{k-\frac{1}{2}}),
\end{split}\label{4.14}
\end{align}
\normalsize which preserves the discrete extended canonical two
form
\begin{align}
\frac{1}{2}(d\p_{k}\wedge d\q_{k+1}+d\p_{k+1}\wedge d\q_{k})
  -\psi^{m}(\q_{k+\frac{1}{2}}, \p_{k+\frac{1}{2}})\wedge
   \left(\frac{dt_{k}+dt_{k+1}}{2}\right). \label{4.15}
\end{align}
The integrator (\ref{4.14}) is a two-step symplectic-energy
integrator with $2m$th order of accuracy.


\sect{Concluding remarks}

We have developed a discrete total variation calculus in
Hamiltonian formalism in this paper. This calculus provides a new
method for constructing structure-preserving integrators for
Hamiltonian systems from a variational view of point. Using this
calculus, we have  derived discrete energy conservation laws
associated to the integrators with variable time-steps. The coupled
integrators are two-step integrators and preserve a discrete
version of the extended canonical two form. If we take fixed
time-steps, the resulting integrators are equivalent to the
symplectic integrators derived directly from the Hamiltonian
systems only in special cases. Thus, some new two-step symplectic
integrators have variationally been obtained. Using generating
function method, we have also obtained high order
symplectic-energy integrators.

In principle, our discussions can be generalized to multisymplectic Hamiltonian
system
\begin{equation}
   M\z_{t}+K\z_{x}=\bigtriangledown_{\z}H(\z),\quad \z\in \mathbf{R}^{n}, \label{5.1}
\end{equation}
where $M$ and $K$ are skew-symmetric matrices on $\mathbf{R}^{n},n\geq 3$ and $
S : \mathbf{R}^{n}\to \mathbf{R}$ is a smooth function [\ref{b1}, \ref{b2}]. We call the above system
  multi-symplectic Hamiltonian system, since it has a multi-symplectic conservation
law
\begin{equation}
    \frac{\partial}{\partial t}\omega +\frac{\partial}{\partial x}\kappa=0,\label{5.2}
\end{equation}
where $\omega $ and $\kappa$ are the pre-symplectic forms
\[
    \omega =\frac{1}{2}d\z\wedge Md\z, \quad  \quad
     \kappa =\frac{1}{2}d\z\wedge Kd\z.
\]

Constructing action functional
\begin{align}
 S=\int\left(\frac{1}{2}\z^{T}(M\z_{t}+K\z_{x})-H(\z)\right)dx\wedge dt \label{5.3}
\end{align}
and performing total variation on (\ref{5.3}), we obtain the
multisymplectic Hamiltonian system (\ref{5.1}) as well as the
corresponding local energy conservation law
\begin{equation}
    \frac{\partial}{\partial t}\left(S(z)-\frac{1}{2}z^{T}Kz_{x}\right)
  +\frac{\partial}{\partial x}\left(\frac{1}{2}z^{T}Kz_{t}\right)=0,
\end{equation}
and the corresponding local momentum conservation law
\begin{equation}
    \frac{\partial}{\partial t}\left(\frac{1}{2}z^{T}Mz_{x}\right)
   +\frac{\partial}{\partial x}\left(S(z)-\frac{1}{2}z^{T}Mz_{t}\right)=0.
\end{equation}

In the same way, we can develop a discrete total variation in
multisymplectic formulation and obtain
multisymplectic-energy-momentum integrators. We shall explore this
issue elsewhere.

On the other hand, we may also get
symplectic/multisymplectic-energy-momentum integrators by means of
the difference variational principle [\ref{glw}], [\ref{guo2}] in
view of regarding difference with variable steps as an entire
object. In this approach, the difference Legendre transformation
may be introduced so as to the discrete total variations in
Lagrangian and Hamiltonian formalism may be transformed to each
other. We shall also explore this issue elsewhere.




\end{document}